\documentclass[pra,showpacs,floatfix,twocolumn]{revtex4}
\usepackage{graphicx}
\usepackage{bm,amsmath}

\newcommand{\fref}[1]{Fig.~\ref{#1}}
\newcommand{\Fref}[1]{Figure \ref{#1}}
\newcommand{\sref}[1]{Sec. \ref{#1}}
\newcommand{\Eref}[1]{Eq.~(\ref{#1})}
\newcommand{\tref}[1]{Table~\ref{#1}}
\def\veps{\varepsilon}

\begin{document}

\title{Parity nonconservation in electron recombination of
multiply charged ions}

\author{G. F. Gribakin}
\email[E-mail address: ]{g.gribakin@am.qub.ac.uk}
\author{F. J. Currell}
\affiliation{Queen's University, Belfast BT7 1NN, Northern Ireland, United
Kingdom}
\author{M. G. Kozlov}
\email[E-mail address: ]{mgk@MF1309.spb.edu}
\author{A. I. Mikhailov}
\affiliation{Petersburg Nuclear Physics Institute, Gatchina 188300, Russia}

\date{\today }

\begin{abstract}
We discuss a parity nonconserving asymmetry in the cross section of
KLL dielectronic recombination of polarized electrons on the
hydrogen-like ions with $Z \lesssim 60$. This effect is strongly
enhanced because of the near-degeneracy of doubly-excited $2l2l'$
states of opposite parity in He-like ions. For ions with $Z \sim 30$
the asymmetry is of the order of $10^{-9}$. For $Z \approx 48$ a
level crossing takes place, leading to the PNC asymmetry of $\pm
5\times 10^{-9}$, which is $10^8$ times greater than the basic
strength of the weak interaction in atoms.
\end{abstract}

\pacs{32.80.Ys,34.80.Lx,11.30.Er}

\maketitle

%------------------------------------------------------------------

%%%%%%%%%%%%%%%%%%%%%%
\section{Introduction}
\label{Intro}%%%%%%%%%

Parity nonconservation (PNC) is caused by the weak interaction. According
to the standard model this interaction is described in terms of charged and
neutral currents. The charged currents play a dominant role in nuclei, e.g.
in $\beta$-decay. The neutral currents lead to the PNC electron-nuclear
interaction and can be observed in atomic experiments \cite{Khr91}. In this
paper we propose that enhanced PNC effects can be seen in electron
recombination of multiply charged ions (MCI).

The first suggestions and estimates of PNC effects in MCI were made in 1974 by
\citet{GL74}. A successful observation of PNC effects in
optical experiments with heavy neutral atoms (see the recent review \cite{GF04}
for references) has renewed the interest in PNC effects in MCI
\cite{SSI89,KLN92,Pin93,Dun96,LNP01,NLL02,MS04}. The obvious advantage of MCI
is the $Z^5$ scaling of the PNC matrix elements with the nuclear charge $Z$, as
opposed to $Z^3$ scaling in neutral atoms \cite{Khr91}.

However, this advantage is usually compensated by larger energy differences
between the levels of opposite parity.
Indeed, PNC effects in atoms and ions appear because of the mixing of the
levels of opposite parity. This mixing leads, for example, to an admixture of
a negative-parity state $\psi _-$ to a positive-parity state $\psi _+$ due to
the parity nonconserving weak interaction $H^{\rm PNC}$,
$\psi_+ + i \eta \psi_-$, as determined by the first-order perturbation
expression
%------------------------------------------------------------------
\begin{equation}
i \eta = \frac{\langle -|H^{\rm PNC}|+\rangle}
{E_+ - E_-  +\frac{i}{2} \Gamma_-}.
\label{i1}
\end{equation}
%------------------------------------------------------------------
The mixing coefficient $\eta $ is real when the level width $\Gamma_-$ is
negligible compared to the level spacing $E_+ - E_-$. In neutral atoms the
valence energies are roughly independent of $Z$, and $\eta$ scales as
$Z^3$. In MCI the level energies $E_\pm $ are proportional to $Z^2$ and a
typical PNC mixing $\eta$ again scales as $Z^3$.

In some special cases the levels of opposite parity in MCI can be anomalously
close. For example, levels of the configurations $1s2s$ and $1s2p$ in
He-like ions cross several times as $Z$ varies \cite{GL74}. Their proximity
leads to a strong enhancement of the PNC effects. At the crossing point
($E_-=E_+$) the maximal size of the mixing parameter is limited
by the level widths and can be estimated as:
%------------------------------------------------------------------
\begin{equation}
\eta \sim \left \langle -|H^{\rm PNC}|+\rangle\right
/\Gamma_\pm .
\label{i2}
\end{equation}
%------------------------------------------------------------------
According to Ref.~\cite{GL74}, a crossing of the $1s2p~^3\!P_1$ and
$1s2s~^1\!S_0$ levels takes place at $Z \approx 32$. Because of the
difference in the total electronic angular momentum, these levels can only
be mixed by the nuclear-spin-dependent (NSD) part of the PNC interaction
\cite{NLL02}. Two opposite-parity levels with the total angular momentum
$J=0$, $1s2s~^1S_0$ and $1s2p~^3P_0$, cross twice at larger $Z$, around 65
and 90 \cite{ALP03}. For such ions one can expect enhanced
nuclear-spin-independent (NSI) PNC effects. In both cases the detection
schemes involve radiative transitions.

In this paper we propose to study PNC mixing in He-like ions by looking at
the parity-violating asymmetry in KLL dielectronic recombination (DR) of
electrons with H-like ions. Here the PNC interaction manifests itself
as a difference between the recombination cross sections for electrons
with positive and negative helicities. The observation of such difference
means a correlation between the spin and momentum of the incident electron
of the form $\bm{\sigma }\cdot \bm{p}$, which does violate parity,
since $\bm{p}$ is a vector and $\bm{\sigma }$ is a pseudovector
($\bm{p}\rightarrow -\bm{p}$, while $\bm{\sigma }\rightarrow \bm{\sigma }$
under spatial inversion).

The PNC interaction in DR mixes the intermediate doubly excited
$2s^2$ and $2s2p$ states of the He-like ion, which decay by the
emission of a photon. In this respect PNC effects in DR are
similar to those in neutron scattering from heavy nuclei. PNC
asymmetries of up to 10\% have been observed in nuclei by tuning
the neutron energy to the $p$-wave compound nuclear resonances.
This enhancement over the typical size of the nuclear weak
interaction ($10^{-7}$) is caused by the proximity of $s$- and
$p$-wave resonances, and by the large ratio of the $s$- to
$p$-wave neutron capture amplitudes (see, e.g., review \cite{FG95}
and references therein).

KLL dielectronic recombination can be observed in experiments with Electron
Beam Ion Traps (EBITs) or ion storage rings (see, e.g., \cite{FJCvol1}).
However, these device do not have a polarized ion target or electron beam, as
is required. Furthermore, the present generation of devices do not achieve the
sensitivity required to observe the PNC effect. Criteria for the sensitivity
requirements are outlined in section~\ref{discuss_sect} from which the
feasibility can be established for any future experimental devices.

The doubly excited $2l2l'$ configurations contain a larger number of closely
spaced levels than the singly excited $1s2l$ configurations. In \sref{Energy}
we calculate the energies of the doubly excited $2l2l'$ states for $10\le
Z\le60$ and identify crossings between levels of opposite parity with $\Delta
J=0$ and 1. We then estimate the widths of the close levels of opposite parity
and the PNC mixing coefficients $\eta$. In \sref{sigma} we evaluate the DR
cross section and PNC asymmetries. The paper concludes with a short feasibility
analysis of PNC measurements in recombination of MCI.

%%%MGK>>>
The main aim of our work is to present the first analysis of PNC effects in
the DR on H-like ions, to obtain a reliable estimate of the size
of the PNC effects and to find the resonances and nuclear charges where these
effects are largest. At the next stage it should be possible to improve
significantly the accuracy of the calculations by using the well developed
theory of the H-like and He-like ions.
%%%MGK<<<

%%%%%%%%%%%%%%%%%%%%%%%%%%%%%%%%%%%%%%%%%%%%%
\section{Energy levels for the $2l2l'$ shell}
\label{Energy}%%%%%%%%%%%%%%%%%%%%%%%%%%%%%%%

The energies of the $2l2l'$ states are determined by
diagonalization of the effective Hamiltonian in the $n=2$
subspace. The eigenstates are obtained as $\sum C_{lj,l'j'}
|lj,l'j'\rangle$, where $lj$ and $l'j'$ define the hydrogen-like
orbitals with $n=2$. The single-electron part of this Hamiltonian
includes hydrogenic Dirac orbital energies and the Lamb shift. The
two-electron part of the Hamiltonian matrix for the configurations
$2s^2$, $2p^2$, and $2s2p$ is taken from Ref. \cite{BGS84}. This
work presents it as a double expansion in parameters $1/Z$ and
$\alpha Z$ and we use three terms of this expansion of order $Z$,
$Z(\alpha Z)^2$, and $Z^0$ (atomic units are used throughout the
paper and $\alpha\approx 1/137$ is the fine structure constant).
%%%MGK>>>
In particular, the first term accounts for the Coulomb interaction
between the electrons. The term $Z(\alpha Z)^2$ accounts for the Breit
interaction and for the relativistic corrections to the wave functions.
The last term ($Z^0$) corresponds to the second order in the Coulomb
interaction.
%%%MGK<<<

%------------------------------------------------------------------
\begin{figure}[tb]
\includegraphics[scale=0.45,angle=0]{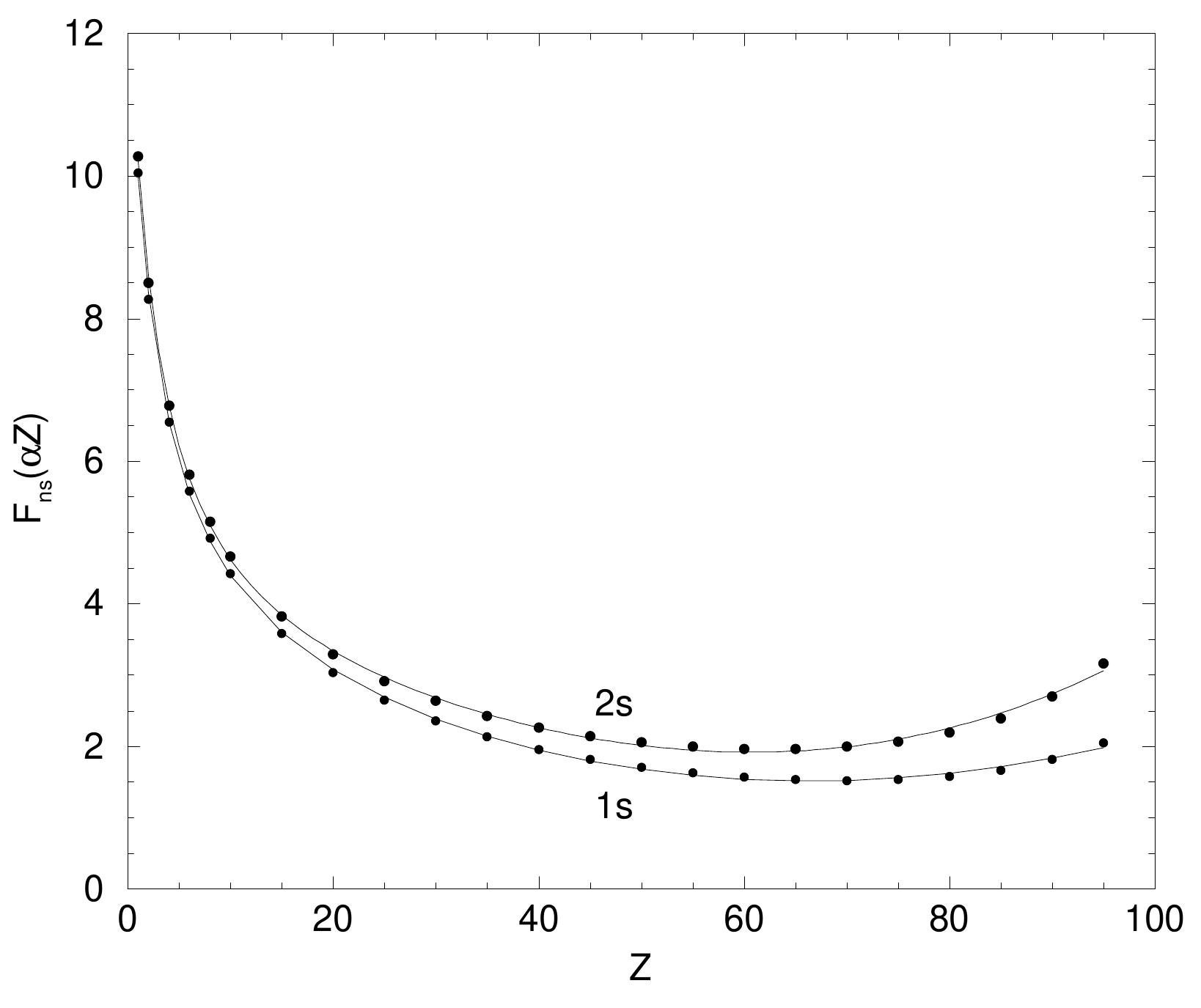}
%\hfill
\includegraphics[scale=0.45]{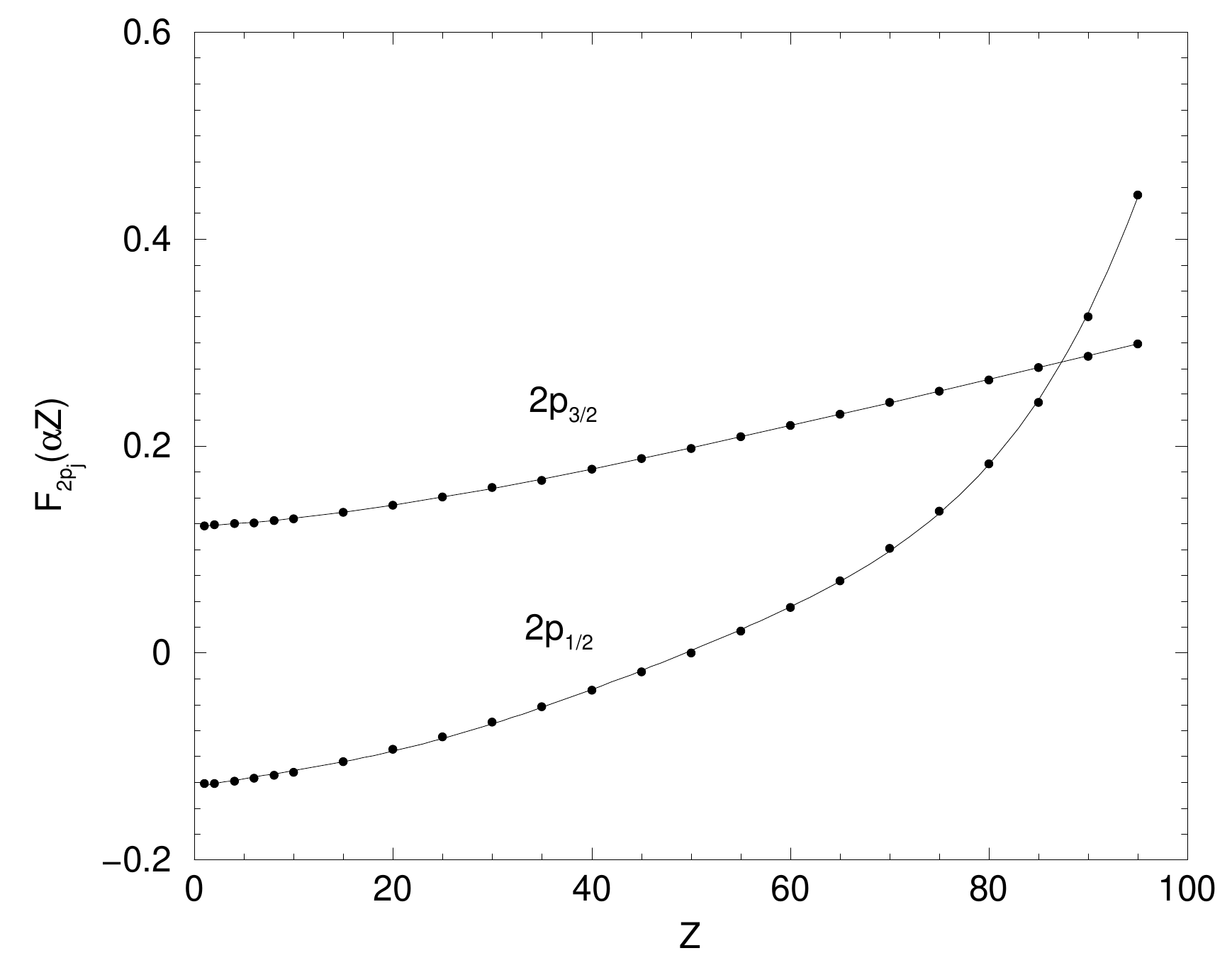}
\caption{The Lamb shift for the $1s$, $2s$, and $2p_j$ orbitals of
the H-like ions. Solid circles show the values of $F_{nlj}$ from
\Eref{e1} calculated by Johnson and Soff \cite{JS85}. The curves
correspond to 5-parameter fits used in our calculations for He-like
ions.}
 \label{fig_lamb}
\end{figure}
%------------------------------------------------------------------

The Lamb shift appears in the higher order, $Z(\alpha Z)^3$, but it is
known to be essential for the level crossings within the $1s2l'$ manifold
\cite{GL74}. Below we show that this is also true for the $2l2l'$ states.
By factoring out the main dependence on $Z$ and the principal quantum number
$n$, the Lamb shift for the hydrogenic orbital $nlj$ is written as
%------------------------------------------------------------------
\begin{equation}
\delta E_{nlj} = \frac{Z(\alpha Z)^3}{\pi n^3} F_{nlj}(\alpha Z).
\label{e1}
\end{equation}
%------------------------------------------------------------------
The values of $F_{nlj}$ calculated for $n=1,2$ and $Z$ up to 95 by
\citet{JS85} are shown in \fref{fig_lamb}. They account for the
self-energy correction, vacuum polarization, and finite nuclear size
effects.

\begin{figure}[t]
\includegraphics[scale=0.48]{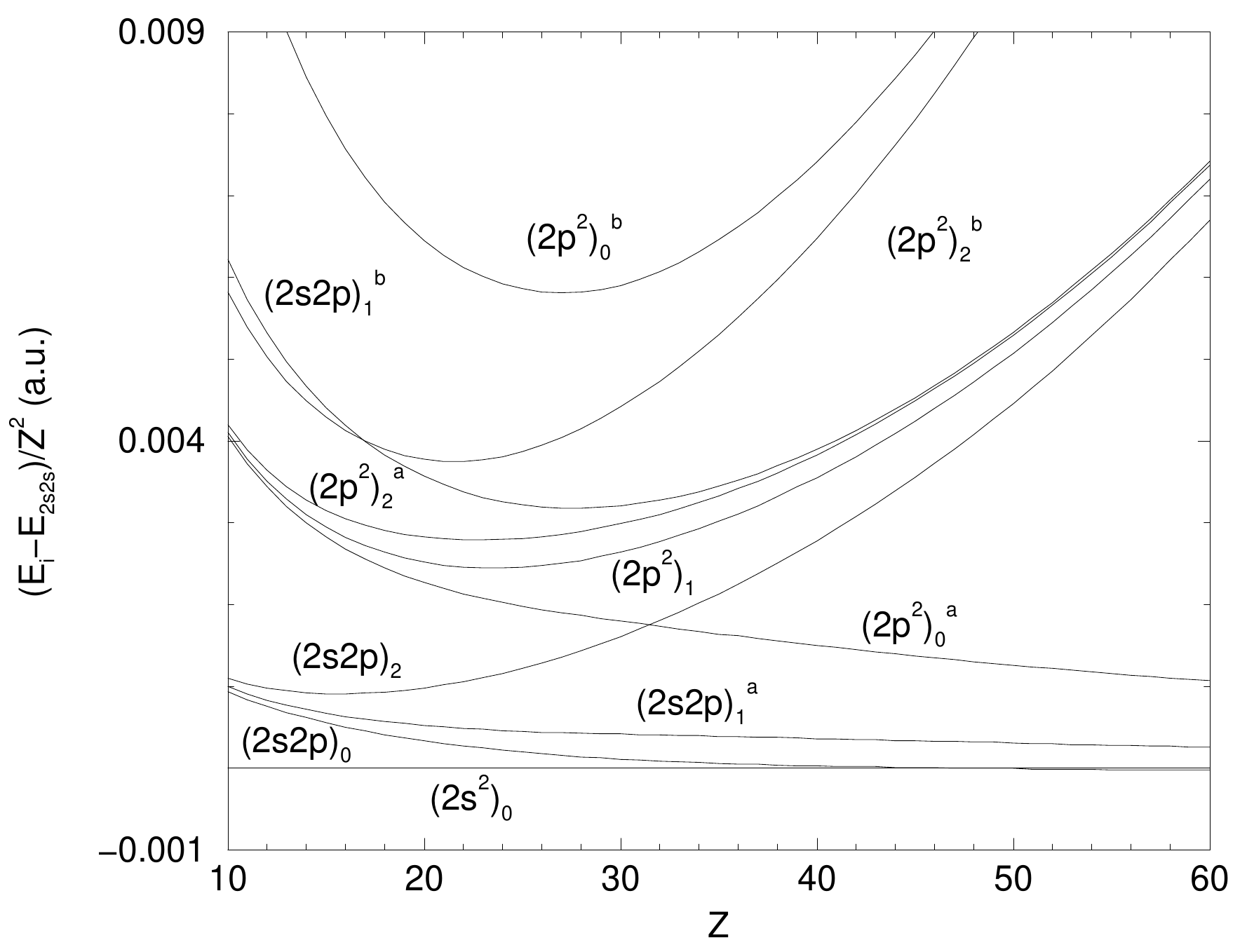}
\caption{Energies of the levels of the $2s2p$ and $2p^2$
configurations relative to the $(2s^2)_0$ level for He-like ions.
All energies are divided by $Z^2$ to account for the general scaling
of energies in MCI.
%%%MGK>>>
The levels are labeled as $(2l2l')_J$ and additional superscripts $a,~b$ are
added to distinguish levels with identical electron configurations and $J$.
%%%MGK<<<
} \label{fig_e2}
\end{figure}

The results of the diagonalization of the Hamiltonian matrix are shown in
\fref{fig_e2}. The eigenstates are labeled as $(2l2l')_J$ and additional
superscripts $a,~b$ are added to distinguish levels with identical quantum
numbers.
One can see two crossings of the levels of opposite parity: a pair of
levels with $\Delta J=1$ cross at $Z \approx 17$ and another pair with
$\Delta J=0$ cross at $Z \approx 48$. The latter crossing between
$(2s^2)_0$ and $(2s2p)_0$ levels is entirely due to the Lamb shift. This
crossing disappears if the Lamb shift is neglected. Instead, another
crossing with $\Delta J=1$ appears near $Z=42$ between the levels
$(2p^2)_2^a$ and $(2s2p)_1^b$. As seen in
\fref{fig_e2}, their energies are very close for $Z>40$.

%------------------------------------------------------------------
%------------------------------------------------------------------

Of course, the exact position of the level crossings may depend on the
higher terms of the $\alpha Z$ expansion, which are neglected here. The
first crossing at $Z\approx 17$ is rather sharp and takes place at
relatively small $Z$, where radiative corrections are still small.
Therefore, this crossing is known rather accurately. The crossing between
the levels $(2s^2)_0$ and $(2s2p)_0$ is more sensitive. However, it should
not disappear when higher-order terms are included because of the $Z^4$
scaling of the Lamb shift \eqref{e1}. Finally, the relative position of the
$(2p^2)_2^a$ and $(2s2p)_1^b$ levels is most sensitive to the higher order
terms and they may still cross at high $Z$.

The crossing at $Z \approx 17$ leads to enhanced NSD PNC effects. The
crossing at $Z \approx 48$ is favorable for the observation of the NSI part
of the PNC interaction. Here much larger PNC effects can be expected due to
their scaling with $Z$. In addition, NSI interaction is roughly $Z$ times
stronger than NSD interaction, see \sref{pnch}. Because of that we focus on
the effects due to PNC mixing of the $(2s^2)_0$ and $(2s2p)_0$ levels.

According to \Eref{i2} the PNC mixing near the crossing points depends on
the line widths. For autoinizing states the total width is the sum of the
radiative and autoionizing widths, $\Gamma=\Gamma^{(r)}+\Gamma^{(a)}$. In
the non-relativistic hydrogenic approximation the radiative widths of the
states $(2s2p)_0 \equiv (-,0)$ and $(2s^2)_0 \equiv (+,0)$ are given by:
%------------------------------------------------------------------
\begin{align}
     \Gamma^{(r)}_{-,0} &= \left(\frac{2}{3}\right)^{\!\!8}
     \alpha^3 Z^4 = 1.517 \times 10^{-8} Z^4,
\label{e2}\\
     \Gamma^{(r)}_{+,0} &= 2\left(\frac{2}{3}\right)^{\!\!8}
     \alpha^3 Z^4 \left(1-C_{ss}^2\right),
\label{e3}
\end{align}
%------------------------------------------------------------------
where the coefficient $C_{ss}$ defines contribution of the configuration
$2s^2$ to the state $|+,0\rangle$:
%------------------------------------------------------------------
\begin{align}
    |+,0\rangle &=
    C_{ss}|2s,2s\rangle + C_{pp}|2p_{1/2},2p_{1/2}\rangle
 \nonumber\\
    &\quad + C_{p'p'}|2p_{3/2},2p_{3/2}\rangle.
 \label{e4}
\end{align}
%------------------------------------------------------------------

The autoionizing widths are evaluated in Appendix~\ref{width}.
They depend weakly on $Z$ and for $Z > 30$ are smaller than the
radiative widths. Hence, for $Z > 30$ the size of the PNC mixing
\eqref{i2} is limited largely by the radiative width, $\Gamma_{\pm,0}
\approx \Gamma^{(r)}_{\pm,0}$. This means that the exact position
of the level crossing is not very important for calculation of the
PNC effect.

%%%%%%%%%%%%%%%%%%%%%%%%
\section{PNC Hamiltonian and mixing}
\label{pnc}%%%%%%%%%%%%%

%%%%%%%%%%%%%%%%%%%%%%%%%%%%%%%%%%%%%%%%%%%%%%%%%%%%%%%%%%%%%%%%
\subsection{PNC Hamiltonian and one-electron PNC matrix element}
\label{pnch}%%%%%%%%%%%%%%%%%%%%%%%%%%%%%%%%%%%%%%%%%%%%%%%%%%%%

The Hamiltonian of the PNC interaction consists of the NSI and NSD parts
and in relativistic notation has the form \cite{Khr91}:
%------------------------------------------------------------------
\begin{align}
     H^{\rm PNC} &= H^{\rm PNC}_{\rm NSI} + H^{\rm PNC}_{\rm NSD}
\nonumber\\
     &= \frac{G_{\rm F}}{\sqrt{2}}
     \Bigl(-\frac{Q_{\rm W}}{2} \gamma_5 + \frac{\kappa}{I}\,
     \gamma_0 \bm{\gamma}\cdot \bm{I}\Bigr) n(\bm{r}),
\label{p1}
\end{align}
%------------------------------------------------------------------
where  $G_{\rm F}=2.2225\times 10^{-14}$~a.u. is the Fermi constant of the
weak interaction, $\gamma_i$ are the Dirac matrices, $\bm{I}$~is the
nuclear spin, and $n(\bm{r})$ is the nuclear density normalized as $\int
n({\bm r}) d {\bm r}=1$. The dimensionless constants $Q_{\rm W}$ and $\kappa$
characterize the strength of the NSI and NSD parts, respectively. $Q_{\rm
W}$ is known as the weak charge of the nucleus. In the lowest order the
standard model yields:
%------------------------------------------------------------------
\begin{equation}
     Q_{\rm W} = -N + Z (1 - 4\sin^2\theta_{\rm W}) \approx - N,
\label{p2}
\end{equation}
%------------------------------------------------------------------
where $N$ is the number of neutrons and $\theta_{\rm W}$ is the Weinberg
angle, $\sin^2 \theta_{\rm W}\approx 0.23$. Radiative corrections to
\Eref{p2} change $Q_{\rm W}$ by few percent \cite{Hag02}.

The constant $\kappa$ includes contributions from the anapole moment
$\kappa_{\rm a}$ and from the electron-nucleon neutral currents
$\kappa_{eN}$ $(|\kappa_{eN}| \ll 1)$. Flambaum and Khriplovich showed
that $\kappa_{\rm a} \sim \alpha A^{2/3}$, where $A=Z+N$ is the number
of nucleons, and for heavy nuclei dominates over the constant
$\kappa_{eN}$ \cite{FK80}. One more contribution to the constant
$\kappa$ was calculated by \citet{FK80} and by \citet{BP91b}. Except for
the very heavy nuclei, this contribution is significantly smaller than
that of the anapole moment.

Weak charges of the nuclei $^{203}$Tl and $^{133}$Cs were measured with
high accuracy by \citet{VMM95} and  by \citet{WBC97}. These measurements
played an important role in low-energy tests of the standard model (see
review \cite{ER04}). Up to now the only measurement of the NSD PNC
amplitude was made for $^{133}$Cs \cite{WBC97}. A detailed discussion of
this matter and a complete list of references can be found in the recent
review \cite{GF04}.

Because of the short-range nature of the interaction in
Hamiltonian \eqref{p1} it effectively mixes only one-electron
states with $j=1/2$, i.e.~$ns_{1/2}$ and $\tilde{n}p_{1/2}$. For a
point-like nucleus the corresponding matrix element turns to
infinity because of the singular behavior of the Dirac orbitals
at the origin. For a finite nucleus of the radius $R_{\rm nuc}$
this matrix element can be approximately given by the following
expression \cite{BB74,Khr91}:
%------------------------------------------------------------------
\begin{align}
\label{me2}
&\langle \tilde{n}p_{1/2}|H^{\rm PNC}|ns_{1/2}\rangle =
\\
&\quad=-i \frac{\sqrt{2}\,G_{\rm F}\alpha Z^4 R}{8\pi(\tilde{n}n)^{3/2}}
\left(Q_W + \frac{4\gamma_{1/2}+2}{3}\,
\frac{2}{I}(\bm{I}\cdot \bm{j})\kappa\right),
\nonumber
\end{align}
%------------------------------------------------------------------
where $R$ is the relativistic enhancement factor:
%------------------------------------------------------------------
\begin{align}
R &=
\frac{4\left(2ZR_{\rm nuc}/r_{\rm Bohr}\right)^{2\gamma_{1/2}-2}}
{\Gamma^2(2\gamma_{1/2}+1)}\,,
\label{me3}\\
\gamma_j &\equiv \left[(j+1/2)^2-(\alpha Z)^2\right]^{1/2},
\label{me3a}
\end{align}
%------------------------------------------------------------------
and the following approximation can be used for the nuclear radius:
%------------------------------------------------------------------
\begin{equation}
     R_{\rm nuc}=1.2\, A^{1/3}\, {\rm Fm}
     = 2.27\times 10^{-5} A^{1/3}.
\label{me1}
\end{equation}
%------------------------------------------------------------------

The accuracy of these expressions is a few percent, at least for
the NSI part. A more accurate calculation can easily be done using
Dirac orbitals for the finite nucleus. At 1\% level of accuracy
the details of the nuclear structure and radiative corrections
become important (see \cite{KF03,MS04,MS02,DP02a} and references
therein).

%%%%%%%%%%%%%%%%%%%%%%%%%%%%%%%%%%%%%%%%%
\subsection{PNC mixings for He-like ions}
\label{pncm}%%%%%%%%%%%%%%%%%%%%%%%%%%%%%

Let us examine the enhancement of the PNC mixing due to the proximity
of the levels of opposite parity in He-like ions. The mixing parameter
$\eta$ is estimated with the help of Eqs. \eqref{i1} and \eqref{me2}.
The NSI part of the PNC Hamiltonian mixes
only states with $\Delta J=0$. \Fref{fig_e2} shows that there is only one
pair of close levels of opposite parity which meets this requirement,
namely $(2s^2)_0$ and $(2s2p)_0$. The first of these states has admixtures
of the configurations $2p^2_{1/2}$ and $2p^2_{3/2}$ [see \Eref{e4}]. For
example, for $Z=32$ the weights of these configurations are 0.19 and 0.02,
respectively. Thus, the interaction between configurations $2s^2$ and
$2p^2$ should be taken into account.
Using Eqs. \eqref{e4} and \eqref{me2}, we obtain:
%------------------------------------------------------------------
\begin{align}
\label{mix2}
&\langle -,0|H^{\rm PNC}_{\rm NSI}|+,0\rangle\!
=\\
&\quad=-i \frac{\sqrt{2}\,G_{\rm F}}{64\pi}\alpha Z^4 R Q_W
\left(C_{ss}-C_{pp}\right)
\equiv i h^{\rm PNC}.
\nonumber
\end{align}
%------------------------------------------------------------------

\begin{table*}[ht]
\caption{PNC mixing $\eta $ between levels $(2s^2)_0$ and $(2s2p)_0$
and comparison of its scaling with typical scaling of the PNC mixing in
neutral atoms $(Z^3R)$.}
\label{tab1}
\begin{ruledtabular}
\begin{tabular}{lrrrrrrrrrrr}
$Z$
%%%%%%%%%%% calculated with program mci_pnc.for:
&$  10  $&$  15  $&$  20  $&$  25  $&$  30  $
&$  35  $&$  40  $&$  45  $&$  50  $&$  55  $&$   60   $\\
\hline
$\Re\, \eta \times 10^{11}$
&$-0.068$ & $-0.40 $ & $-1.59 $ & $ -6.20$ & $-19.5 $
&$-58.2 $ & $ -159 $ & $  -204 $ & $  150  $ & $ 383  $ & $ 531    $\\
$\Im\, \eta \times 10^{11}$
&$ 0.004$ & $ 0.018 $ & $ 0.076 $ & $ 0.41 $ & $  2.2 $
&$ 13.0 $ & $  84.3 $ & $  417 $ & $ 637  $ & $ 620  $ & $ 642    $\\
$\dfrac{|\eta |}{Z^3R}\times 10^{15}$
&$ 0.65 $&$ 1.08 $&$ 1.69 $&$ 3.11 $&$  5.17$
&$ 8.85 $&$ 15.8 $&$ 24.9 $&$ 22.0 $&$ 15.6 $&$ 11.4   $\\
\end{tabular}
\end{ruledtabular}

\end{table*}

Results of the calculation of the PNC mixing for different $Z$ are
presented in \tref{tab1}. The resonant enhancement at the level crossing
is not very pronounced. Firstly, the level crossing is not sharp. Secondly,
for $Z>30$ the radiative width, which grows as $Z^4$, becomes greater
than the autoionizing width and for $Z>40$ it exceeds the level spacing.
As a result, the absolute value of the PNC mixing grows steadily with $Z$.
However, the enhancement at the level crossing is clearly seen when we
consider the mixing strength divided by the $Z^3R$ PNC scaling parameter.
The real part of the mixing changes sign at the resonance, where $\eta $
is equal to:
%------------------------------------------------------------------
\begin{equation}
\eta_{Z=48} = (0 +6.0\, i) \times 10^{-9}.
\label{mix2b}
\end{equation}
%------------------------------------------------------------------

%%%%%%%%%%%%%%%%%%%%%%%%%%%%%%%%%%%%%%%%%%%%%%%%%%%%%%%%%%%%%%%
\section{PNC effect in Dielectronic Recombination}
\label{sigma}%%%%%%%%%%%%%%%%%%%%%%%%%%%
%%%%%%%%%%%%%%%%%%%%%%%%%%%%%%%%%%%%%%%%%%%%%%%%%%%%%%%%%%%%%%%

\begin{figure}[tb]
\includegraphics[scale=0.75]{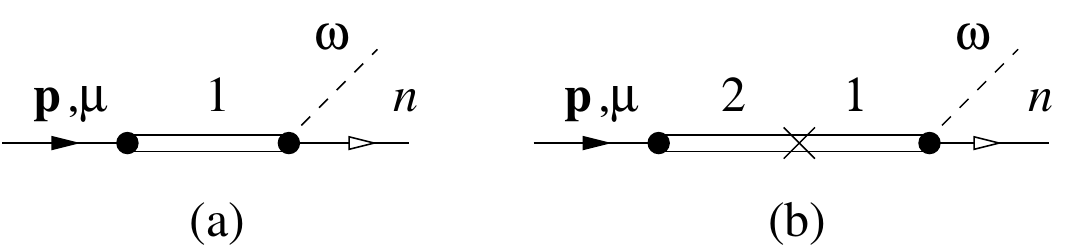}
\caption{Diagrammatic representation of the DR amplitude \eqref{t1}:
(a) is the conventional contribution and (b) is the PNC correction.
The initial state $\bm{p},\mu$ describes the electron with momentum
$\bm{p}$ and helicity $\mu=\bm{\sigma }\cdot \bm{p}/2p = \pm 1/2$
incident on the H-like ion in the $1s$ ground state. The double
lines 1 and 2 correspond to the doubly excited states of the He-like
ion which decay by emission of a photon $\omega$ to the final state
$n$. The cross denotes PNC mixing of the states 1 and 2.}
\label{fig_sigma}
\end{figure}
%------------------------------------------------------------------

The formalism we use to calculate the PNC asymmetry in DR is similar
to that developed for PNC effects in neutron scattering (see, e.g.,
review by \citet{FG95}). DR is
described by the diagrams in \fref{fig_sigma}, where we assume
that the incident electron $|\bm{p},\mu\rangle$ has the energy
$\veps$, which is close to the transition energy between the
ground state of the H-like ion and the levels $|i=1,2\rangle$ of
the configuration $2l_i2l_i'$ of the corresponding He-like ion.
Then the contribution of resonance 1 to the DR amplitude $A$ is:
%\begin{widetext}
%------------------------------------------------------------------
{\fontsize{10pt}{\baselineskip}
\begin{align}
  &
  A \equiv A^{\rm PC} + A^{\rm PNC}
     =\frac{i \sqrt{{2\pi\omega}/{V}}
     \langle n |\bm{e}_q \cdot \bm{r}|1\rangle}
     {E_{1s}+\veps- E_1 + \frac{i}{2} \Gamma_1}
\label{t1} \\
 &\!
     \left\{\!\!\langle 1 |V_{\rm C}| \bm{p},\mu; 1s_{1/2,M}\rangle
%     \right.
%     &&\left.
     \!+\!\frac{\langle 1 |H^{\rm PNC}| 2\rangle
     \langle 2 |V_{\rm C}| \bm{p},\mu; 1s_{1/2,M}\rangle}
     {E_{1s}+\veps- E_2 + \frac{i}{2}\Gamma_2}\!\!\right\}\!,
 \nonumber
 \end{align}}
%------------------------------------------------------------------
 \noindent
where $\bm{e}_q$ and $\omega$ define the polarization and frequency
of the photon, $1s_{1/2,M}$ describes the initial state of the
target with spin projection $M$, and $V_{\rm C}$ is the Coulomb
interaction. We use dipole approximation for radiative transition
and $V$ is the quantization volume for the electromagnetic field
\cite{Sob79}. The total width of the doubly excited states 1 and 2
is given by the sum of the radiative and autoionizing widths,
$\Gamma_i=\Gamma_i^{(r)}+\Gamma_i^{(a)}$.

As we have seen in \sref{pncm}, the strongest PNC effect is expected
for the crossing of two levels with $J_1=J_2=0$, which simplifies
the derivation. Hence we assume that the states 1 and 2 in
\fref{fig_sigma} correspond to the levels
$|\pm,0\rangle$ discussed in \sref{pncm}.

The total DR cross section is given by the sum over the final
states of the ion and polarization of the photon:
%------------------------------------------------------------------
\begin{align}
  \sigma &\equiv \sigma^{\rm PC}\! + \sigma^{\rm PNC}\!
     = \frac{2\pi}{p}\sum_{q,n}\!\int\!
     \left\{\!\left|A^{\rm PC}\right|^2\!
     + 2\Re \left(A^{\rm PC\,*} A^{\rm PNC}\right)\!\right\}
  \nonumber\\
  &\times
     \delta(E_{1s}+\veps-\omega-E_n) Vd\rho_\omega,
\label{t2}
\end{align}
%------------------------------------------------------------------
where $d\rho_\omega=\omega^2d\omega d\Omega/(2\pi c)^3$
and we neglect the square of the small PNC amplitude.
When we substitute \Eref{t1} in \eqref{t2}, both PC and PNC terms
appear to be proportional to the radiative width $\Gamma^{(r)}_1$:
%------------------------------------------------------------------
\begin{align}
     \Gamma^{(r)}_1 &= 2\pi\sum_{q,n}
     \left|i\sqrt{2\pi\omega/V}\langle 1 |\bm{e}_q \cdot \bm{r}|n\rangle
     \right|^2
   \nonumber\\
     &\times\int\delta(E_{1s}+\veps-\omega-E_n) Vd\rho_\omega
   \nonumber\\
     &=\frac{4\omega^3\sum_n |\langle n||r|| 1 \rangle|^2}{3c^3(2J_1+1)}.
\label{t3}
\end{align}
%\end{widetext}
%------------------------------------------------------------------
The last expression is standard (see \cite{Sob79})
and was used  to calculate $\Gamma^{(r)}_{\pm,0}$ in Eqs. \eqref{e2}
and \eqref{e3}. The PC cross section now reads:
%------------------------------------------------------------------
\begin{eqnarray}
     \sigma^{\rm PC}_1
     &=& \frac{1}{p}\,
     \frac{\Gamma^{(r)}_1
     |\langle \bm{p},\mu; 1s_{1/2,M} |V_{\rm C}|1\rangle|^2}
     {\left(E_{1s}+\veps- E_1\right)^2 + \frac14\Gamma_1^2}.
\label{t4}
\end{eqnarray}
%------------------------------------------------------------------
The remaining Coulomb matrix element determines the autoionizing width
$\Gamma^{(a)}_1$:
%------------------------------------------------------------------
\begin{align}
\label{t5}
  \Gamma^{(a)}_1 &=
     \frac{p}{\pi(2J_1+1)}\sum_{M_1,M,\mu}
     |\langle \bm{p},\mu; 1s_{1/2,M} |V_{\rm C}|1\rangle|^2
\\
     &=\frac{2p}{\pi}\sum_{M}
     |\langle \bm{p},\mu; 1s_{1/2,M} |V_{\rm C}|1\rangle|^2,
\nonumber
\end{align}
%------------------------------------------------------------------
where we take into account that $J_1=0$. Note that the sum in \eqref{t5}
does not depend on the electron helicity $\mu$, or on the direction of
its momentum $\bm{p}$, while the individual matrix elements do depend on
$\mu$ and $M$. Introducing the branching ratio
$R_{\mu,M}^{(a)}$ for autoionization into channel
$(\mu,M)$, we can rewrite \eqref{t4} in the final form:
%------------------------------------------------------------------
\begin{eqnarray}
     \sigma^{\rm PC}_1
     &=& \frac{\pi}{2p^2}\,
     \frac{\Gamma^{(r)}_1 \Gamma^{(a)}_1 R_{\mu,M}^{(a)}}
     {\left(E_{1s}+\veps- E_1\right)^2 + \frac14\Gamma_1^2}.
\label{t6}
\end{eqnarray}
%------------------------------------------------------------------

Similarly, the PNC contribution to the cross section becomes:
%------------------------------------------------------------------
\begin{widetext}
\begin{align}
  \sigma^{\rm PNC}_1
  \!\! =\!
     \frac{2}{p}
     \Re\!\left\{\!\!
     \frac{\Gamma^{(r)}_1\!
     \langle \bm{p},\mu; 1s_{1/2,M} |V_{\rm C}|1\rangle
     \langle 1 |H^{\rm PNC}| 2\rangle
     \langle 2 |V_{\rm C}| \bm{p},\mu; 1s_{1/2,M}\rangle}
     {\left(E_{1s}+\veps- E_2 + \frac{i}{2}\Gamma_2\right)
     \left(\left(E_{1s}+\veps- E_1\right)^2 + \frac14\Gamma_1^2\right)}
     \!\!\right\}
%\nonumber\\
%  &=
     \!\!=\!
     2 \sigma^{\rm PC}_1 \Re\!\left\{\!\!
     \frac{\langle 2 |V_{\rm C}| \bm{p},\mu; 1s_{1/2,M}\rangle
     \langle 1 |H^{\rm PNC}| 2\rangle}
     {\langle 1 |V_{\rm C}|\bm{p},\mu; 1s_{1/2,M}\rangle
     \!\left(E_{1s}+\veps- E_2 + \frac{i}{2} \Gamma_2\right)}
     \!\!\right\}\!.
\label{t7}
\end{align}
\end{widetext}
%------------------------------------------------------------------
Further simplification of \Eref{t7} requires an explicit form of the Coulomb
matrix elements. Let us expand the incident electron state in partial waves,
%------------------------------------------------------------------
\begin{align}
     &|\bm{p}, \mu\rangle\! =\!
     {\frac{(2\pi)^{3/2}}{\sqrt{p}}}\!
     \sum_{j,l,m}
     \!\langle\Omega_{j,l,m}(\hat{\bm{p}})
     |\chi_\mu(\hat{\bm{p}})\rangle\,
     i^{l} e ^{i \delta_{jl}}
     |\veps,j,l,m\rangle,
\label{t8}
\end{align}
%------------------------------------------------------------------
where $\Omega_{j,l,m}$ and $\chi_\mu $ are spherical and ordinary spinors
and $\delta_{jl}$ is the scattering phase shift.
Wave function \eqref{t8} is normalized so that
$\langle\bm{p}', \mu'|\bm{p}, \mu\rangle
= (2\pi)^3\delta(\bm{p}'-\bm{p})\delta_{\mu',\mu}$,
and the radial functions are normalized to the delta function of energy,
$\langle\veps',j',l',m'|\veps,j,l,m\rangle
= \delta(\veps'-\veps)\delta_{j',j}\delta_{l',l}\delta_{m',m}$.
If we direct the quantization axis along $\hat{\bm{p}}$, the spinor matrix
element in \eqref{t8} can be written explicitly:
%------------------------------------------------------------------
\begin{align}
  \langle\Omega_{j,l,m}(\hat{\bm{p}})
     |\chi_\mu(\hat{\bm{p}})\rangle
     &= \sum_\lambda C^{j,m}_{l,\lambda,1/2,\mu}
     Y^*_{l,\lambda}(\hat{\bm{p}})
     \nonumber\\
     &= C^{j,\mu}_{l,0,1/2,\mu}
     \left(\frac{2l+1}{4\pi}\right)^{1/2}\!
     \delta_{m,\mu}\,,
\label{t8a}
\end{align}
%------------------------------------------------------------------
where $C^{j,m}_{l,\lambda,s,\mu}$ is the Clebsh-Gordon coefficient and
$Y_{l,\lambda}$ is the spherical harmonic.

When we use expansion \eqref{t8} to calculate the Coulomb matrix elements in
\Eref{t7}, the angular and parity selection rules leave only one term of
this expansion with $j=\frac12$ and $l=l_i=0$, or 1 depending on the parity
$P_i$ of the intermediate state $i$: $l_i=(1-P_i)/2$:
%------------------------------------------------------------------
\begin{align}
  \langle i |V_{\rm C}|\bm{p},\mu; 1s_{1/2,M}\rangle
     &= \frac{(2\pi)^{3/2}}{\sqrt{p}}
     \langle\Omega_{1/2,l_i,-M}(\hat{\bm{p}})
     |\chi_\mu(\hat{\bm{p}})\rangle
   \nonumber\\
     &\times
     i^{l_i} e ^{i \delta_i}
     \langle P_i,0|V_{\rm C}|\veps,\tfrac12,l_i,\!-M; 1s_{1/2,M}\rangle ,
\label{t9}
\end{align}
%------------------------------------------------------------------
where $\delta _i\equiv \delta _{1/2,l_i}$.
Substituting \eqref{t8a} and \eqref{t9} in \eqref{t5} we obtain the following
expression for the autoionizing width:
%------------------------------------------------------------------
\begin{eqnarray}
     \Gamma^{(a)}_i =
     4\pi |\langle P_i,0 |V_{\rm C}|\veps,\tfrac12,l_i,\mu;
     1s_{1/2,-\mu}\rangle|^2.
\label{t9a}
\end{eqnarray}
%------------------------------------------------------------------
We can also use \eqref{t8a} to find the branching ratio
$R_{\mu,M}^{(a)}$ in \eqref{t6}:
%------------------------------------------------------------------
\begin{eqnarray}
     R_{\mu,M}^{(a)} = \delta_{\mu,-M}.
\label{t9b}
\end{eqnarray}
%------------------------------------------------------------------
This expression is valid only if the quantization axis for the
angular momentum of the target ion coincides with the direction of
the momentum of the incident electron. Averaging over
polarizations of the beam and the target gives $\langle
R_{\mu,M}^{(a)}\rangle=1/2$ and \Eref{t6} transforms into the
standard Breit-Wigner expression \cite{LL77}:
%------------------------------------------------------------------
\begin{eqnarray}
     \sigma^{\rm PC}_1
     &=& \frac{\pi}{4p^2}\,
     \frac{\Gamma^{(r)}_1 \Gamma^{(a)}_1}
     {\left(E_{1s}+\veps- E_1\right)^2 + \frac14\Gamma_1^2}.
\label{t6a}
\end{eqnarray}
%------------------------------------------------------------------

Intermediate levels 1 and 2 in \Eref{t7} have different parity leading
to different partial wave contributing to the matrix element \eqref{t9}:
$l_1=1-l_2$. The  corresponding spherical
spinors are related by (see, e.g., Ref.~\cite{Khr91}):
%------------------------------------------------------------------
\begin{eqnarray}
     \Omega_{1/2,l_2,m}(\hat{\bm{p}})
     = -(\bm{\sigma}\cdot \hat{\bm{p}})
     \Omega_{1/2,l_1,m}(\hat{\bm{p}}),
\label{t10}
\end{eqnarray}
%------------------------------------------------------------------
where $(\bm{\sigma}\cdot \hat{\bm{p}})\chi_\mu(\hat{\bm{p}})
=2\mu \chi_\mu(\hat{\bm{p}})$. The partial matrix elements in \eqref{t9}
are real and we can write:
%%------------------------------------------------------------------
%\begin{align}
%  &\frac{\langle 2 |V_{\rm C}| \bm{p},\mu; 1s_{1/2,-\mu }\rangle}
%     {\langle 1 |V_{\rm C}|\bm{p},\mu; 1s_{1/2,-\mu }\rangle}
%     = i^{l_1-l_2} e ^{i(\delta_1-\delta_2)}
%     (\bm{\sigma}\cdot \hat{\bm{p}})
%     \nonumber\\
%   &\quad \times \frac{\langle P_2,0 |V_{\rm C}|\veps,\frac12,l_2,\mu ;
%     1s_{1/2,-\mu }\rangle}
%     {\langle P_1,0 |V_{\rm C}|\veps,\frac12,l_1,\mu ; 1s_{1/2,-\mu }\rangle}
%\label{t11}\\
%  &\quad\quad =
%     i\,e ^{i(\delta_1-\delta_2)}\eta_{1,2}
%     (\bm{\sigma}\cdot \hat{\bm{p}})
%     \sqrt{{\Gamma_2^{(a)}}/{\Gamma_1^{(a)}}}.
%\label{t12}
%\end{align}
%%------------------------------------------------------------------
\begin{align}
  &\frac{\langle 2 |V_{\rm C}| \bm{p},\mu; 1s_{1/2,-\mu }\rangle}
     {\langle 1 |V_{\rm C}|\bm{p},\mu; 1s_{1/2,-\mu }\rangle}
     = -i^{l_2-l_1} e^{i(\delta_2-\delta_1)}
     (\bm{\sigma}\cdot \hat{\bm{p}})
     \nonumber\\
   &\quad \times \frac{\langle P_2,0 |V_{\rm C}|\veps,\frac12,l_2,\mu ;
     1s_{1/2,-\mu }\rangle}
     {\langle P_1,0 |V_{\rm C}|\veps,\frac12,l_1,\mu ; 1s_{1/2,-\mu }\rangle}
\label{t11}\\
  &\quad\quad =
     -i^{l_2-l_1} e ^{i(\delta_2-\delta_1)}\eta_{1,2}
     (\bm{\sigma}\cdot \hat{\bm{p}})
     \sqrt{{\Gamma_2^{(a)}}/{\Gamma_1^{(a)}}}.
\label{t12}
\end{align}
The factor $\eta_{1,2}=\pm 1$ in \eqref{t12} depends on the signs of the
partial matrix elements in \eqref{t11}. Therefore, the PNC cross
section \eqref{t7} takes the form:
%------------------------------------------------------------------
%\begin{eqnarray}
%     \sigma^{\rm PNC}_1
%     &=& 2\eta_{1,2} (\bm{\sigma}\cdot \hat{\bm{p}})\,\sigma^{\rm PC}_1
%     \left(\frac{\Gamma_2^{(a)}}{\Gamma_1^{(a)}}\right)^{1/2}
%     \nonumber\\
%     &\times& \Re\left\{e ^{i(\delta_1-\delta_2)}
%     \frac{i \langle 1 |H^{\rm PNC}| 2\rangle}
%     {E_{1s}+\veps- E_2 + \frac{i}{2} \Gamma_2}
%     \right\}.
%\label{t13}
%\end{eqnarray}
%------------------------------------------------------------------
\begin{eqnarray}
     \sigma^{\rm PNC}_1
     &=& -2\eta_{1,2} (\bm{\sigma}\cdot \hat{\bm{p}})\,\sigma^{\rm PC}_1
     \left(\frac{\Gamma_2^{(a)}}{\Gamma_1^{(a)}}\right)^{1/2}
     \nonumber\\
     &\times& \Re\left\{e ^{i(\delta_2-\delta_1)}
     \frac{i^{l_2-l_1} \langle 1 |H^{\rm PNC}| 2\rangle}
     {E_{1s}+\veps- E_2 + \frac{i}{2} \Gamma_2}
     \right\}.
\label{t13}
\end{eqnarray}

\Eref{t13} is valid for the polarized as well as unpolarized
target. In the first case one should use \Eref{t6}, while in the
second case \Eref{t6a} applies. For the unpolarized electron beam
$\sigma^{\rm PNC}$ must be averaged over the helicity and
\Eref{t13} gives zero for the unpolarized target. However, for the
polarized target $\sigma^{\rm PNC}$ is not zero, because
$\sigma^{\rm PC}$ in \Eref{t6} selects the helicity through the
branching ratio \eqref{t9b}. In fact we can substitute
$(\bm{\sigma}\cdot \hat{\bm{p}})$ in \eqref{t13} with
$-2(\bm{S}\cdot \hat{\bm{p}})$, where $\bm{S}$ is the spin of the
ion.
%------------------------------------------------------------------

%%%%%%%%%%%%%%%%%%%%%%%%%%%%%%%%%%%%%%%%%%%%%%%%%%%%%%%%%%%%%%%%%
\section{Results}
\label{pncr}%%%%%%%%%%%%%%%%%%%%%%%%%%%%%%%%%%%%%%%%%%%%%%%%%%%%%

Now we apply the formalism developed in the previous sections to
calculate the PNC effect in the DR cross section at the energies
near the $(\pm,0)$ resonances in the He-like ions. In the diagrams
in \fref{fig_sigma} and in corresponding equations \eqref{t6a} and
\eqref{t13} the states 1 and 2 can be either $(2s^2)_0$ and
$(2s2p)_0$, or vice versa. These two contributions lead to the
final states $n$ with different parities and we sum the
corresponding cross sections,
$\sigma^{\rm PC}=\sigma^{\rm PC}_1 +\sigma^{\rm PC}_2$.

Equations \eqref{w3}, \eqref{w5}, and \eqref{w6} show that the phase
factor $\eta_{1,2}$ in \eqref{t13} is equal to 1. The incident
electron energy scales as $Z^2$ and the Coulomb phase shifts in
the non-relativistic approximation are independent of $Z$,
$\delta_{sp} \equiv\delta_s-\delta_p \approx 0.953$. Taking this
into account and using \eqref{t6a} and \eqref{t13}, we obtain the
following total PNC cross section:
%------------------------------------------------------------------
%\begin{widetext}
%{\fontsize{10pt}{\baselineskip}
%\begin{align}
%     \sigma^{\rm PNC}
%     &=- \frac{\pi(\bm{\sigma}\cdot\hat{\bm{p}})
%     \sqrt{\Gamma_+^{(a)}\Gamma_-^{(a)}} h^{\rm PNC}}
%     {2p^2 \left(\Delta_+^2+\frac14\Gamma_+^2\right)
%     \left(\Delta_-^2+\frac14\Gamma_-^2\right)}
%\label{pr1}
%%\\ &\times
%     \left[\!\left(\Gamma^{(r)}_+\Delta_-\! -\Gamma^{(r)}_-\Delta_+\!\right)
%     \cos\delta_{sp}\!
%     +\tfrac12\left(\Gamma^{(r)}_+\Gamma_-\! +\Gamma^{(r)}_-\Gamma_+\!\right)
%     \sin\delta_{sp}
%     \right],
%%\nonumber
%\end{align}}
%\end{widetext}
%------------------------------------------------------------------
\begin{align}
     \sigma^{\rm PNC}
     =&- \frac{\pi(\bm{\sigma}\cdot\hat{\bm{p}})
     \sqrt{\Gamma_+^{(a)}\Gamma_-^{(a)}} h^{\rm PNC}}
     {2p^2 \left(\Delta_+^2+\frac14\Gamma_+^2\right)
     \left(\Delta_-^2+\frac14\Gamma_-^2\right)}
 \nonumber \\
     &\times
     \left[\!\left(\Gamma^{(r)}_+\Delta_-\! +\Gamma^{(r)}_-\Delta_+\!\right)
     \cos\delta_{sp}
     \right.
 \nonumber \\
     &-\left.
     \tfrac12\left(\Gamma^{(r)}_+\Gamma_-\! -\Gamma^{(r)}_-\Gamma_+\!\right)
     \sin\delta_{sp}
     \right],
\label{pr1}
\end{align}
%------------------------------------------------------------------
where $\Delta_\pm \equiv E_{1s}+\veps-E_\pm$, $h^{\rm PNC}$ is given by
\eqref{mix2}, and again we can substitute $(\bm{\sigma}\cdot
\hat{\bm{p}})$ with $-2(\bm{S}\cdot \hat{\bm{p}})$ for a polarized
target rather than a polarized electron beam.

%------------------------------------------------------------------
\begin{figure*}[tbh]
\includegraphics[scale=0.45]{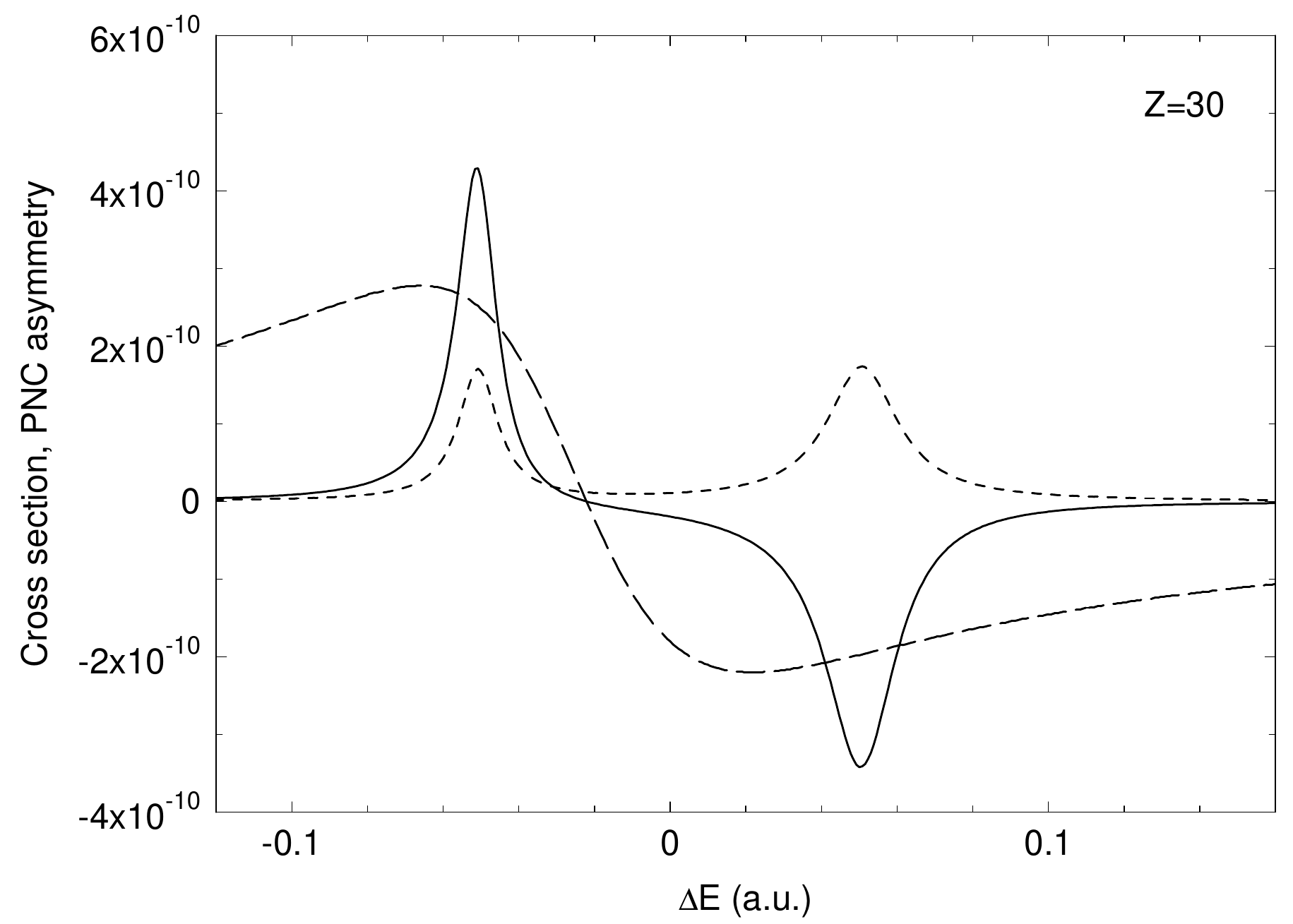}
\hfill
\includegraphics[scale=0.45]{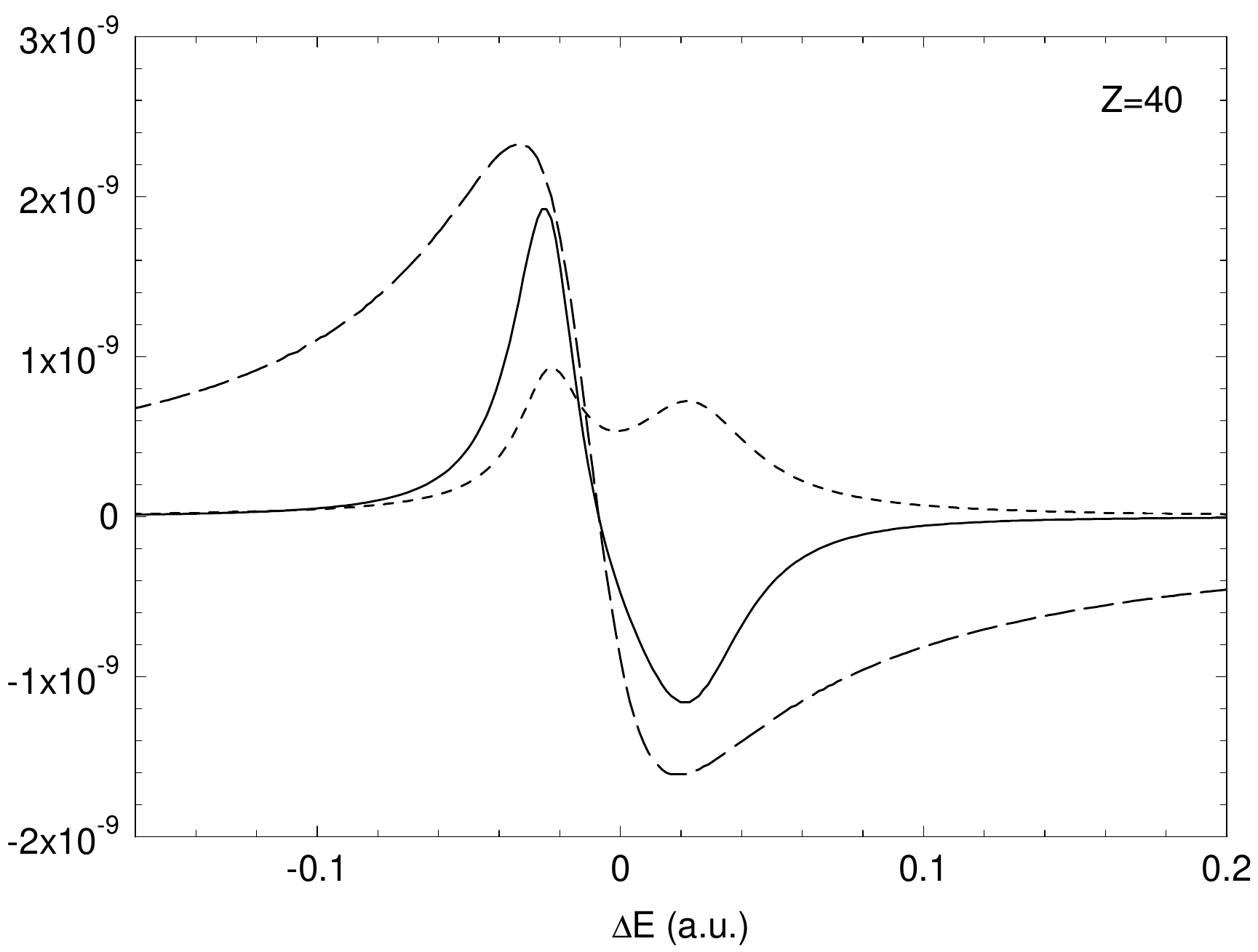}

\vspace{2mm}
\includegraphics[scale=0.45]{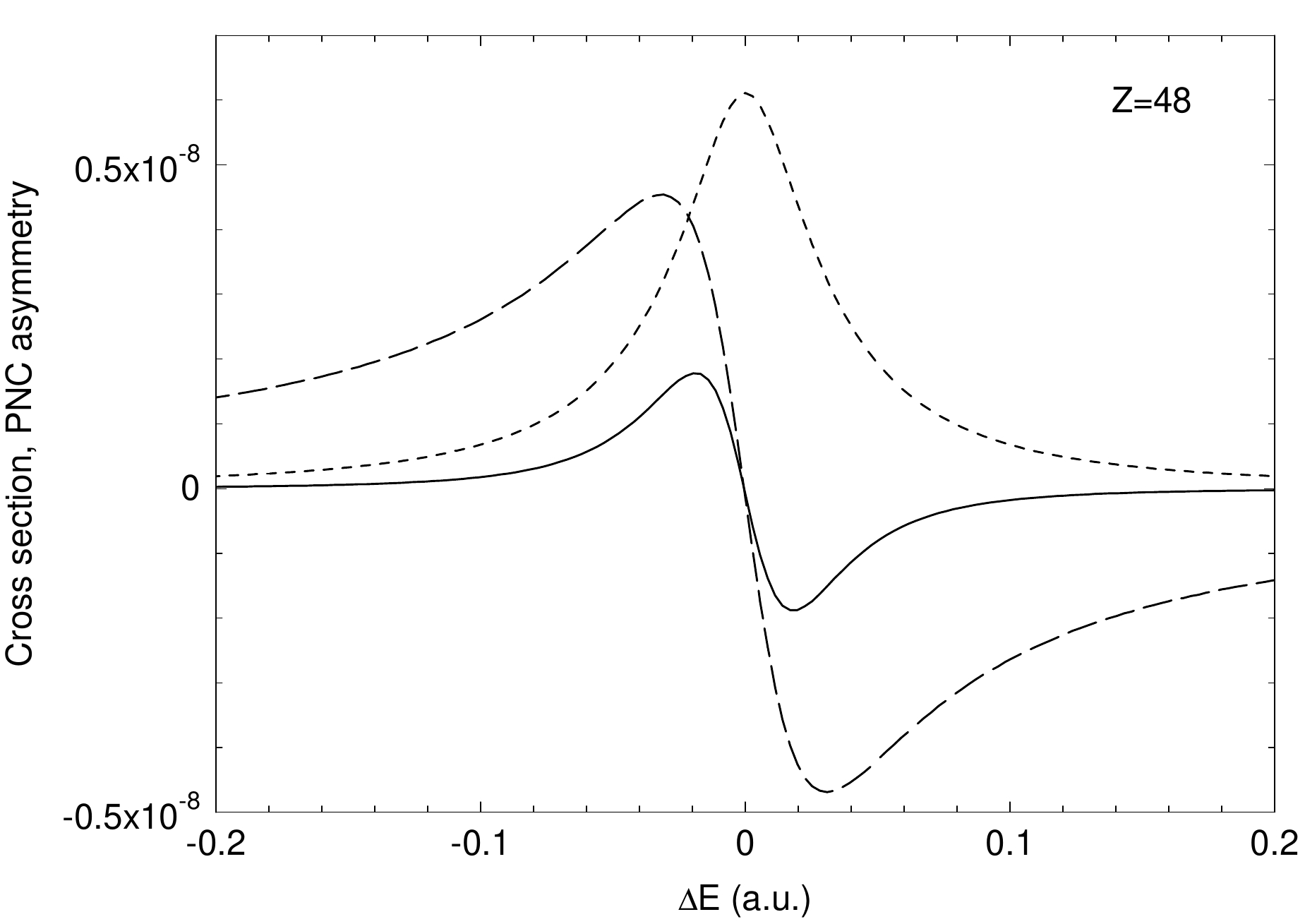}
\hfill
\includegraphics[scale=0.45]{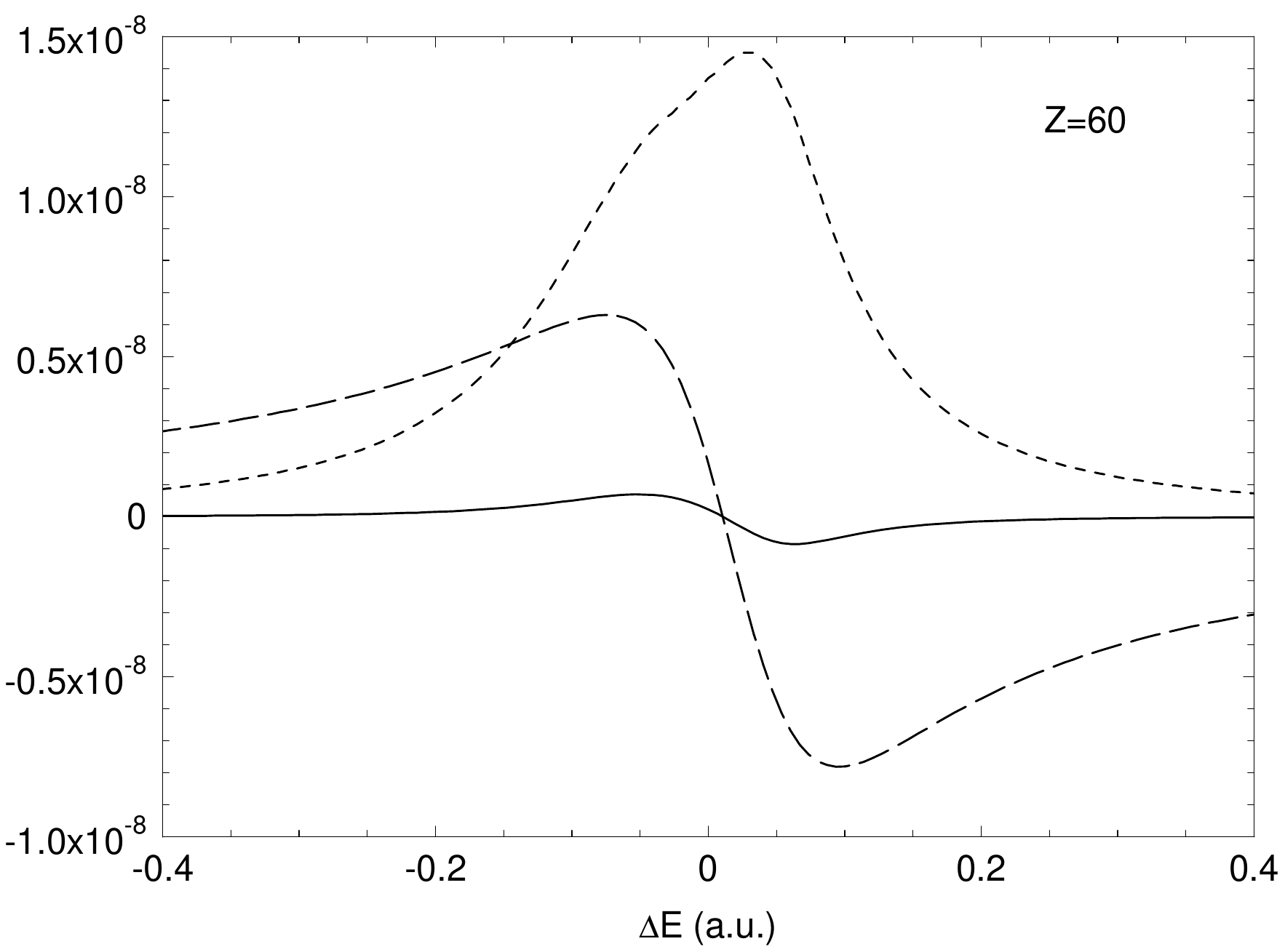}
\caption{PC and PNC DR cross sections (in a.u.) and PNC asymmetry
for $(2s^2)_0$ and $(2s2p)_0$ resonances in H-like ions with $Z=$
30, 40, 48, and 60. The energy $\Delta E=\veps-E_{\rm av}+Z^2/2$,
where $\veps$ and $-Z^2/2$ are the energies of the incident electron
and H-like target and $E_{\rm av}=(E_{(2s^2)_0}+E_{(2s2p)_0})/2$.
Solid lines correspond to $10^3\times\sigma^{\rm
PNC}|_{\bm{\sigma}\cdot\hat{\bm{p}}=1}$,
%%%MGK>>>
long-dashed
%%%MGK<<<
lines are the PNC asymmetry
$\cal A$, and
%%%MGK>>>
short-dashed
%%%MGK<<<
lines correspond to $10^{-n}\times \sigma^{\rm PC}$, where
$n=7,6,5,4$ for $Z=30,40,48,60$.} \label{fig_pnc}
\end{figure*}
%------------------------------------------------------------------

\Fref{fig_pnc} presents the plots of $\sigma^{\rm PC}$,
$\sigma^{\rm PNC}$ and the PNC asymmetry,
%------------------------------------------------------------------
\begin{equation}
     {\cal A} = \frac{\sigma^{+}-\sigma^{-}}{\sigma^{+}+\sigma^{-}}
     \simeq\left. \frac{\sigma^{\rm PNC}}
     {\sigma^{\rm PC}}\right|_{\bm{\sigma}\cdot\hat{\bm{p}}=1},
\label{pr2}
\end{equation}
%------------------------------------------------------------------
where $\sigma^\pm$ are the cross sections for positive and negative
helicity. The peak magnitude of the asymmetry increases from
$3\times 10^{-10}$ for $Z=30$ to $5\times 10^{-9}$ for $Z=48$, i.e.
at the crossing point. It practically does not grow at larger $Z$.
\Fref{fig_pnc} also shows that for $Z\geq 40$ the two resonances
overlap.

For $Z>30$ the radiative width dominates over the autoionizing width
and second term in square brackets in \eqref{pr1} is suppressed. The
first term changes sign between the resonances. Therefore, the PNC
cross section and asymmetry also change sign. Consequently, the net
asymmetry integrated over energy is suppressed. Growth of the PNC
matrix element for higher $Z$ is compensated by the increase in the
width and PNC asymmetry stops growing.

%Comparison of \fref{fig_pnc} with \tref{tab1} shows that the amplitude of the
%asymmetry $\cal A$ is of the order of the PNC mixing $\eta$. For $Z=30$, where
%the imaginary part of $\eta$ is small, the first term in brackets in \Eref{pr1}
%dominates. \Fref{fig_e2} shows that for $Z<48$, the level $(+,0)$ lies below
%the level $(-,0)$. As a result, the PNC cross section \eqref{pr1} and the
%asymmetry $\cal A$ are mostly positive. For $Z=40$, where the real and
%imaginary parts of $\eta$ are close, the second term in brackets in \eqref{pr1}
%becomes important. As a result, the asymmetry changes sign through the
%resonance. Finally, for $Z=48$ and $Z=60$ the imaginary part of $\eta$ is
%larger and the asymmetry is mostly negative. Note, that after the level
%crossing, the two terms in \eqref{pr1} add constructively between the
%resonances. This leads to a single negative peak for the PNC asymmetry for
%$Z=60$.

%%%%%%%%%%%%%%%%%%%%%%%%%%%%%%%%%%%%%%%%%%%%%%%%%%%%%%%%%%%%%%
\section{Discussion and conclusions}
\label{discuss_sect}
%%%%%%%%%%%%%%%%%%%%%%%%%%%%%%%%%%%%%%%%%%%%%%%%%%%%%%%%%%%%%%

It is useful to estimate the feasibility of measuring the PNC asymmetry in KLL
recombination calculated above and from this estimate derive the sensitivity
requirements for an experimental apparatus capable of observing the PNC using
the scheme proposed. The number of counts in an experiment with a fully
polarized electron beam with positive helicity is given by:
\begin{equation}
\label{d1}
N_+= j_e N_i t \epsilon\sigma^+ \equiv I\sigma^+,
\end{equation}
where $j_e$ is the electron flux, $N_i$ is the number of target
ions, $t$ is the acquisition time, and $\epsilon$ is the detection
efficiency. The number of counts for negative helicity is
$N_-=I\sigma^-$.

For a beam or target with polarization $P$, to detect the PNC asymmetry, the
difference between the counts, $P |N_+ -N_-|$ should be greater than
statistical error, $\sqrt{N_+ +N_-}$, which gives:
\begin{equation}
\label{d2}
  I > \frac{\sigma^+ + \sigma^- + 2\sigma_b}{P^2(\sigma^+ - \sigma^-)^2},
\end{equation}
where $\sigma_b$ is the magnitude of any background occurring through direct
radiative recombination or as an experimental artifact (e.g. detector dark
counts). We express this as a cross section for convenience although for
experimental artifact background signal, this is the effective cross section to
which the apparatus background would correspond. For the rest of this analysis
we consider the ideal limit, $P=1$, $\sigma_b = 0$.

Equation \eqref{d2} is valid for a mono-energetic electron beam. If
the electron energy spread in the beam is greater than the
resonance spacing and widths, then the flux $j_e$ in \eqref{d1}
should be replaced by the flux density $dj_e/d\varepsilon$. The
counts $N_{\pm}$ are obtained by integrating over the electron
energy and the effect can be detected if
\begin{equation}
\label{d3}
  I_{\rm av} > \int(\sigma^+ + \sigma^-) d \veps
  \left(\int (\sigma^+ - \sigma^-) d \veps\right)^{-2}.
\end{equation}
The first integral above is equal to $2(S_1+S_2)$, where
\begin{equation}
\label{d4}
  S_i = \frac{\pi^2}{2p^2}\,
  \frac{\Gamma_i^{(r)}\Gamma_i^{(a)}}{\Gamma_i},
\end{equation}
is the strength of resonance $i$. The integral
$\int (\sigma^+ - \sigma^-) d \veps$ in
\Eref{d3} can be written as $2S_{1,2}^{\rm PNC}$, where
%\begin{align}
%  S_{1,2}^{\rm PNC} \equiv&
%  \int \left.\sigma^{\rm PNC}\right|_{\bm{\sigma}\cdot\hat{\bm{p}}=1} d \veps
%\label{d5}\\
%  =& -\frac{\pi^2}{p^2}\,
%     \frac{\sqrt{\Gamma_+^{(a)}\Gamma_-^{(a)}} h^{\rm PNC}
%     \left(\frac{\Gamma_+^{(r)}}{\Gamma_+}+\frac{\Gamma_-^{(r)}}{\Gamma_-}\right)}
%     {\left[(E_+-E_-)^2+\frac14(\Gamma_+ + \Gamma_-)^2\right]}
%\nonumber\\
%     &\times\left[(E_+-E_-)\cos\delta_{sp}
%     +\tfrac12(\Gamma_++\Gamma_-)
%     \sin\delta_{sp} \right],
%  \nonumber
%\end{align}
\begin{align}
  S_{1,2}^{\rm PNC} \equiv&
  \int \left.\sigma^{\rm PNC}\right|_{\bm{\sigma}\cdot\hat{\bm{p}}=1} d \veps
\label{d5}\\
  =& -\frac{\pi^2}{p^2}\,
     \frac{\sqrt{\Gamma_+^{(a)}\Gamma_-^{(a)}} h^{\rm PNC}
     \left(\frac{\Gamma_+^{(r)}}{\Gamma_+}
          -\frac{\Gamma_-^{(r)}}{\Gamma_-}\right)}
     {\left[(E_+-E_-)^2+\frac14(\Gamma_+ + \Gamma_-)^2\right]}
\nonumber\\
     &\times\left[(E_+-E_-)\cos\delta_{sp}
     -\tfrac12(\Gamma_++\Gamma_-)
     \sin\delta_{sp} \right],
  \nonumber
\end{align}
is the PNC strength of the two resonances. Thus, \Eref{d3} reads:
\begin{equation}\label{d6}
  I_{\rm av} > \tfrac12(S_1+S_2)/\left(S_{1,2}^{\rm PNC}\right)^2.
\end{equation}

\begin{figure}[tbh]
\includegraphics*[scale=0.45]{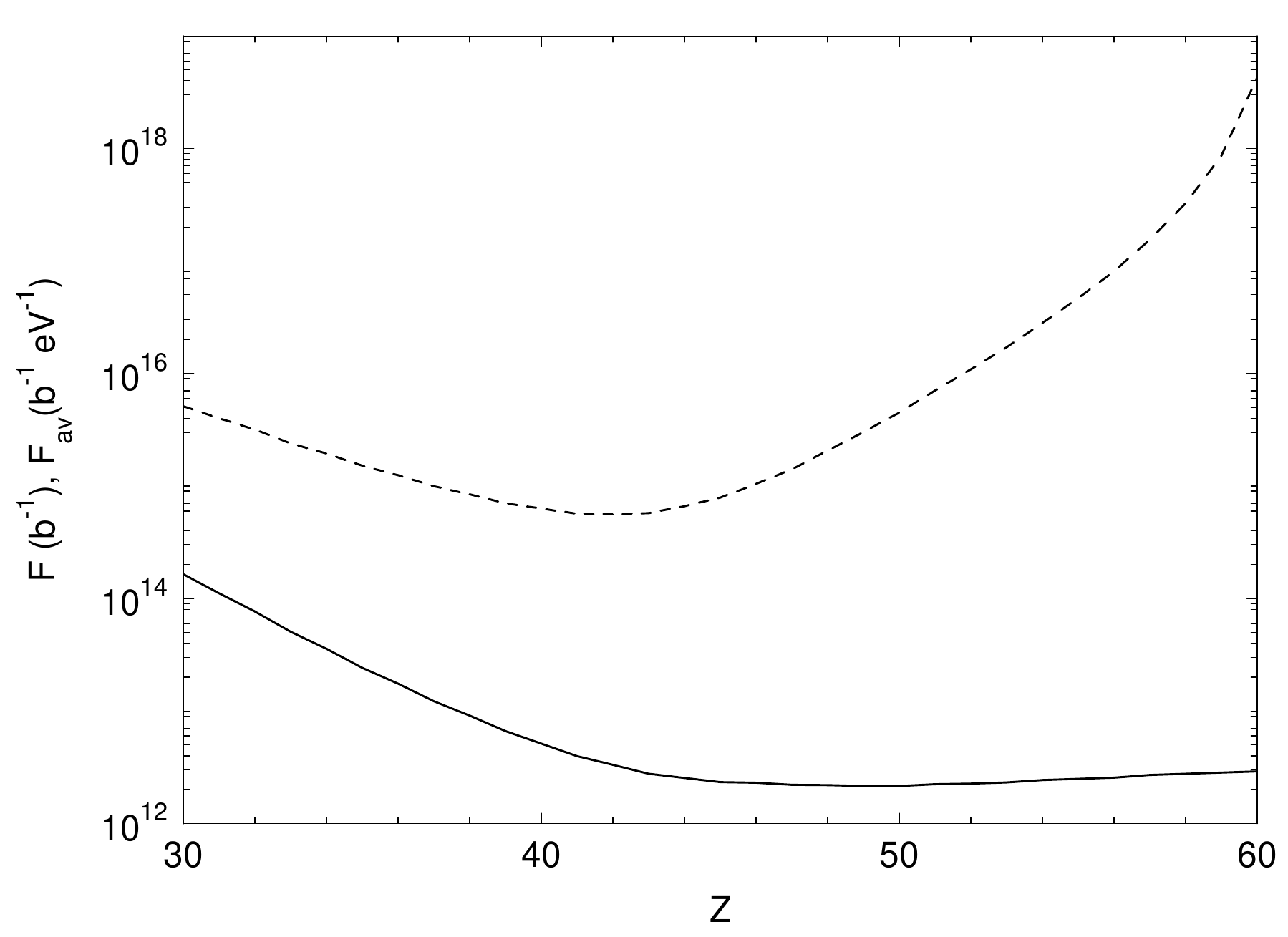}
\caption{PNC measurement feasibility function $F$ in b$^{-1}$ (solid
line) and $F_{\rm av}$ in b$^{-1}$eV$^{-1}$ (dashed line). Shallow
minimum of $F$ at $Z\approx 45$ approximately corresponds to the
level crossing.} \label{fig_fred}
\end{figure}

Equations \eqref{d2} and \eqref{d6} show that for the two limiting cases of
narrow and wide energy distribution in the beam the feasibility of the
experiment on ions with nuclear charge $Z$ depends on the functions:
\begin{align}
F&=
{\rm min}\left[(\sigma^+ + \sigma^-)/(\sigma^+ - \sigma^-)^2\right],
\label{d7}\\
F_{\rm av}
&=
\int(\sigma^+ + \sigma^-) d \veps
\left[\int (\sigma^+ - \sigma^-) d \veps\right]^{-2}\!\!,
\label{d8}
\end{align}
where minimum is taken with respect to the energy of the beam. These functions
are shown in \fref{fig_fred}, where cross sections are in barns
(10$^{-24}$~cm$^2$) and energy in eV.
%As expected, measurements become more
%feasible at large $Z$, where the PNC asymmetry is greater. We also see that the
%averaged PNC effect is strongly suppressed for ions with $38\le Z\le 40$ and
%then steadily grows with $Z$ ($F_{\rm av}$ decreases). This suppression is
%caused by the change of the sign of the asymmetry through the resonances (see
%\fref{fig_pnc}) and it is absent for the mono-energetic electron beam. In the
%latter case the strongest effect can be seen for $Z=50$, i.e. near the level
%crossing.

For mono-energetic beam the measurements become more feasible at
large $Z$, where the PNC asymmetry is greater. The strongest effect
can be seen for $Z\approx 45$, i.e. near the level crossing. The
averaged PNC effect is strongly suppressed by the factor
$(\Gamma_+^{(r)}/\Gamma_+ -\Gamma_-^{(r)}/\Gamma_-)$ in \Eref{d5},
which monotonically decreases with $Z$, as
$\Gamma^{(r)}_\pm/\Gamma_\pm\rightarrow 1$. This suppression is
caused by almost antisymmetric shape of the PNC signal along the
resonances. We conclude that observation of the PNC effects in DR is
much easier with mono-energetic beam.

%------------------------------------------------------------------
%------------------------------------------------------------------

It is worth noting that $F^{-1}$ and $F_{\rm av}^{-1}$ have the same
dimensions as a cross section and a resonance strength respectively.
Indeed, these quantities are useful for estimating the feasibility
of any future experiment designed to observe the PNC effect in KLL
dielectronic recombination resonances. For an experiment to be able
to observe the effect predicted, it would have to be able to detect
a cross section as small as $F^{-1}$ or a resonance strength as
small as $F_{\rm av}^{-1}$, i.e. about $10^{-12}$ b or $10^{-15}$ b
eV, in the absence of background. Of course, this is extremely
demanding, but it is worth remarking that the level crossing
considered here, gives rise to an enhancement of eight orders of
magnitude compared to the basic strength of the weak interaction in
atoms.

Let us compare the present scheme with other proposals for measuring PNC in
ions. \citet{Pin93} suggested to observe PNC effect in the Auger emission from
the He-like uranium. He considered the mixing of the same states, $2s^2$ and
$2s2p$ with $J=0$, and obtained asymmetries of about $10^{-7}$, which is
comparable to our results. However, that estimate neglected the radiative
widths of the levels, which for $Z\gtrsim 50$ exceed the level spacing.

Other proposals were based on the observations of PNC asymmetries in
radiative transitions in He-like ions. \citet{SSI89} focused on the
two-photon $E1$-$M1$ transition between two metastable levels, $2\
{}^3\!P_0 \rightarrow 2\ {}^1\!S_0$, separated by 1~eV in U$^{90+}$.
They showed that the PNC mixing is $|\eta|\sim 5\times 10^{-6}$, and
concluded that lasers with intensities above $10^{21}$ W/cm$^2$
would be required to observe it. In Refs. \cite{KLN92,Dun96,LNP01}
two-photon and hyperfine-quenched transitions $2\ {}^1\!S_0
\rightarrow 1\ {}^1\!S_0$ were examined. Here the mixing between $2\
{}^1\!S_0$ and $2\ {}^3\!P_0$ levels leads to circular polarization
of the photons (up to $10^{-4}$) or to an asymmetry in the photon
angular distribution ($4\times 10^{-4}$ for Gd$^{62+}$ for a fully
polarized ion beam). Although these values seem large, there is a
number of associated problems: low counting rates for the highly
forbidden transitions involved, photon background, detection of the
circular polarization of gamma quanta, and creation of the polarized
ion beam. As a result, the number of events necessary to measure the
effect is large, e.g. $\sim 10^{18}$ \cite{NLL02}.

%------------------------------------------------------------------
\section*{Acknowledgments}
%------------------------------------------------------------------

%%%
Authors are thankful to V.~Shabaev, who pointed at the sign error in
the numerator of \Eref{d5} in the published version of this paper
\cite{GCK05}. In this corrected version of the e-print we traced the
error back to \Eref{t12} and made respective changes to the text and
figures.
%%%

Authors are grateful to A.~Korol, L.~Labzowsky, and A.~Nefiodov for
valuable discussions and to D.~Budker for reading the manuscript.
This work is supported in part by Russian Foundation for Basic
Research, grant No.~05-02-16914. One of us (MK) thanks International
Research Centre for Experimental Physics in Belfast for the DVF
fellowship.

%------------------------------------------------------------------
\appendix

%%%%%%%%%%%%%%%%%%%%%%%%%%%%%%%%%%%%%%%%%%%%%%%%%%%%%%%%%%%%%%%%%
\section{Calculation of autoionizing widths $\Gamma^{(a)}_{\pm,0}$}%%%%%%%
\label{width}%%%%%%%%%%%%%%%%%%%%%%%%%%%%%%%%%%%%%%%%%%%%%%%%%%%%

The autoionizing widths of the doubly excited states with $J=0$
are given by \Eref{t9a}. Their wave functions are linear
combinations of the two-electron states of the form:
%------------------------------------------------------------------
\begin{align}
     &|P,0\, (l_b,l_c)\rangle\!
     =\! \sum_m\! \frac{(-1)^{j-m}} {\sqrt{2j+1}}\,
     |2,l_b,j,m\rangle\, |2,l_c,j,-m\rangle,
 \label{w1b}
\end{align}
%------------------------------------------------------------------
where parity $P=(-1)^{l_b+l_c}$. In the initial state the incident
electron is described by the wave function \eqref{t8} and the
H-like ion is in the ground state $|1s_{1/2,M}\rangle$. The
Coulomb matrix elements on the right-hand-side of \eqref{t9}, after
substituting \eqref{w1b}, are reduced to the two-electron matrix
elements:
%------------------------------------------------------------------
%\begin{widetext}
\begin{align}
 &\langle 2l_b,j_b,m_b;\, 2l_c,j_c,m_c|
 V_{\rm C}|\veps,l_i,j_i,m_i;\,1s,\tfrac12,m \rangle
 \nonumber\\
 &= (-1)^{m_c+m_i+1} [j_b,j_c,j_i,\tfrac12]
 \sum_K \left(\!\begin{array}{ccc} j_i & j_b & K \\
     -m_i & m_b & Q \\ \end{array}\! \right)\!
 \label{w2}\\
 &\times
 \left(\!\begin{array}{ccc} j_c & \tfrac12 & K \\
     -m_c & m & Q \\ \end{array}\! \right)\!
 \left(\!\begin{array}{ccc} j_i & j_b & K \\
     \tfrac12 & -\tfrac12 & 0 \\ \end{array}\! \right)\!
 \left(\!\begin{array}{ccc} j_c & \tfrac12 & K \\
     \tfrac12 & -\tfrac12 & 0 \\ \end{array}\! \right)\!
     R^K_{b,c,i,1s},
\nonumber
\end{align}
%------------------------------------------------------------------
where $[j_a\dots] \equiv [(2j_a+1)\dots]^{1/2}$ and
$R^K_{a,b,c,d}$ is the Coulomb radial integral. It is nonzero for
even $K+l_a+l_c$ and $K+l_b+l_d$. Equations \eqref{w1b} and
\eqref{w2} allow one to calculate the matrix elements in \eqref{t9}.
Neglecting the dependence of the radial integrals on $j$ we have
for the odd state,
%------------------------------------------------------------------
\begin{multline}
  \langle -1,0 |V_{\rm C}|\veps,\tfrac12,l_i,-m; 1s_{1/2,M}\rangle\!
  \\
     = \delta_{l_i,1} \frac{\tau_m}{\sqrt{2}}
     \left(\! R^0_{2p,2s,\veps p,1s}
     -\tfrac{1}{3} R^1_{2s,2p,\veps p,1s}\! \right)\!,
\label{w3}
\end{multline}
%------------------------------------------------------------------
where $\tau_m \equiv (-1)^{m+1/2}$. For the even state \eqref{w1b}
we obtain
%------------------------------------------------------------------
\begin{multline}
  \langle +1,0\, (2l^2) |V_{\rm C}|\veps,\tfrac12,l_i,-m; 1s_{1/2,M}\rangle
 \\
     = \delta_{l_i,0} (-1)^{l} \tau_m \frac{(j+1/2)^{1/2}}{2l+1}
     R^l_{2l,2l,\veps s,1s},
\label{w4}
\end{multline}
%------------------------------------------------------------------
and for the eigenstate \eqref{e4} we arrive at
%------------------------------------------------------------------
\begin{align}
  &\langle +1,0 |V_{\rm C}|\veps,\tfrac12,0,-m; 1s_{1/2,M}\rangle
\label{w5}
 \\
     &= \tau_m \left(\!
     C_{ss}  R^0_{2s,2s,\veps s,1s}
     -\frac{C_{pp}+\sqrt{2}C_{p'p'}}{3}
     R^1_{2p,2p,\veps s,1s}\!\right)\!.
 \nonumber
\end{align}
%\end{widetext}
%------------------------------------------------------------------
%------------------------------------------------------------------

To estimate the widths $\Gamma^{(a)}_i$ we use non-relativistic
hydrogenic radial Coulomb integrals, which do not depend on $Z$:
%------------------------------------------------------------------
\begin{equation}
\begin{array}{cccc}
   R^0_{2s,2s,\veps s,1s} & R^0_{2p,2s,\veps p,1s} &
   R^1_{2s,2p,\veps p,1s} & R^1_{2p,2p,\veps s,1s} \\
            0.0200        &         -0.0304        &
            0.0310        &         -0.0300
\end{array}
\label{w6}
\end{equation}
%------------------------------------------------------------------
Equations \eqref{t9a}, \eqref{w3}, and \eqref{w6} give
%------------------------------------------------------------------
\begin{equation}
     \Gamma^{(a)}_{-,0} = 0.0104,
\label{w7}
\end{equation}
%------------------------------------------------------------------
for all $Z$.\ $\Gamma^{(a)}_{+,0}$ is somewhat smaller than
$\Gamma^{(a)}_{-,0}$ and weakly depends on $Z$ via the coefficients
$C_{aa}$ in \Eref{w5}. For pure $(2s^2)_{0,0}$ state Eqs.\
\eqref{t9a} and \eqref{w4} give $\Gamma^{(a)}_{2s^2} = 0.00496$.
This value is in agreement with Ref.\ \cite{AMM98}.

The same nonrelativistic hydrogenic approximation for the
radiative transitions was used in Eqs.~\eqref{e2} and \eqref{e3} for
the radiative widths $\Gamma^{(r)}_{\pm,0}$.
Again the negative parity state has larger width. Comparison of
Eqs.~\eqref{w7} and \eqref{e2} shows that the radiative width
becomes equal to the autoionizing width for $Z \approx 29$ and
dominates near the level crossing at $Z \approx 48$.

%---------------------------------------------------------------------
%\bibliographystyle{apsrev}
%\bibliography{../bib/julia_w,../bib/my_ref_w}
%\end{document}
%---------------------------------------------------------------------

\end{document}